\pdfoutput=1
\documentclass[conference]{IEEEtran}

\usepackage[utf8]{inputenc}
\usepackage[T1]{fontenc}
\usepackage{newtxtext,newtxmath} % Times-like, crisp in PDF
\usepackage{amsmath,amsfonts}
\usepackage{graphicx}
\usepackage{booktabs}
\usepackage{multirow}
\usepackage{threeparttable}
\usepackage{cite}
\usepackage[hidelinks]{hyperref}
\usepackage{microtype}
\usepackage{url}

\title{Mixture of Low-Rank Adapter Experts in Generalizable Audio Deepfake Detection}

\author{\IEEEauthorblockN{Janne Laakkonen\IEEEauthorrefmark{1}, Ivan Kukanov\IEEEauthorrefmark{2}, and Ville Hautam\"{a}ki\IEEEauthorrefmark{1}}
\IEEEauthorblockA{\IEEEauthorrefmark{1}University of Eastern Finland, Joensuu, Finland\\Email: janne.laakkonen@uef.fi}
\IEEEauthorblockA{\IEEEauthorrefmark{2}KLASS Engineering and Solutions, Singapore}}

\begin{document}
\maketitle

\begin{abstract}
Foundation models such as Wav2Vec2 excel at representation learning in speech tasks, including audio deepfake detection. However, after being fine-tuned on a fixed set of bonafide and spoofed audio clips, they often fail to generalize to novel deepfake methods not represented in training. To address this, we propose a mixture-of-LoRA-experts approach that integrates multiple low-rank adapters (LoRA) into the model's attention layers. A routing mechanism selectively activates specialized experts, enhancing adaptability to evolving deepfake attacks. Experimental results show that our method outperforms standard fine-tuning in both in-domain and out-of-domain scenarios, reducing equal error rates relative to baseline models. Notably, our best MoE-LoRA model lowers the average out-of-domain EER from 8.55\% to 6.08\%, demonstrating its effectiveness in achieving generalizable audio deepfake detection.
\end{abstract}

\section{Introduction}
Significant advances in speech synthesis technology have enabled Text-to-Speech (TTS)~\cite{zhang2023surveyaudiodiffusionmodels} and Voice Conversion (VC)~\cite{Li2022FreevcTH} systems to produce audio indistinguishable from genuine human speech. Malicious actors can exploit synthetic speech to deceive Automatic Speaker Verification (ASV) systems~\cite{Todisco2024MalacopulaAA} or commit fraud~\cite{robinsearly2024}, thereby reducing trust in voice-based authentication platforms. Furthermore, the technology can be used to spread misinformation or impersonate public figures in political and social discourse. Ongoing community efforts, such as ASVspoof challenges~\cite{wu2017asvspoof, todisco2019asvspoof, yamagishi2021asvspoof, wang2024asvspoof}, underscore that \emph{audio deepfake detection} (ADD), often termed speech anti-spoofing, has become a significant research focus. Although notable progress has been made~\cite{li2024audio}, detection models must generalize effectively to out-of-domain or previously unseen attack types. This is a fundamental requirement given the continuous evolution of deepfake generation methods and the difficulty of generalizing detection models across diverse real-world acoustic conditions.

ADD aims to distinguish between genuine (bonafide) and artificially generated (spoofed) audio. Early studies often relied on handcrafted acoustic features such as LFCCs~\cite{alegre2013} and CQCCs~\cite{tak20_odyssey}, but recent efforts have shifted toward self-supervised learning (SSL) frameworks, including Wav2Vec2~\cite{NEURIPS2020_92d1e1eb} and HuBERT~\cite{hsu2021hubert}, which can learn generalized acoustic representations from large-scale unlabeled data. Beyond general representations, Graph Neural Networks (GNNs) \cite{Scarselli2009TheGN} have also shown promising results in ADD, modeling complex relationships between different parts of the audio signal. The spectrotemporal graph attention network AASIST~\cite{jung2022aasist}, designed to capture local spoofing artifacts, has become an effective GNN-based architecture. Tak et al.~\cite{tak22_odyssey} were the first to combine Wav2Vec2 and AASIST, achieving strong results in in-domain evaluations. Despite this progress, current approaches often exhibit a notable performance decrease when faced with unseen attacks or novel acoustic conditions~\cite{muller24b_interspeech, kukanov2024meta}, highlighting a key vulnerability: the reliance on fixed, domain-specific cues, which allows more sophisticated or out-of-distribution spoofing to slip past detection.

Parameter-efficient and adaptive fine-tuning strategies offer a promising approach to address generalization challenges in audio deepfake detection. Techniques such as Low-Rank Adapters (LoRA)~\cite{hu2022lora} and Mixture-of-Experts (MoE)~\cite{jacobs1991, ShazeerMMDLHD17} have shown promise in adapting large pre-trained models to new tasks or domains with limited data. LoRA achieves this by updating only a small subset of model parameters, while MoE dynamically combines the output of multiple specialized ``expert'' networks. Recent work has explored applying these techniques to audio deepfake detection, with promising results \cite{kukanov2024meta, laakkonen2025generalizablespeechdeepfakedetection, zhang2023adaptivefakeaudiodetection, Wang2023LowrankAM}. For instance,~\cite{Wang2023LowrankAM, Wu2024} have demonstrated the effectiveness of applying adapters in Wav2Vec2 for improved performance, while~\cite{Negroni2024} introduced a MoE-based architecture for enhanced generalization across datasets.

\begin{figure*}[t]
    \centering
    \includegraphics[width=\textwidth]{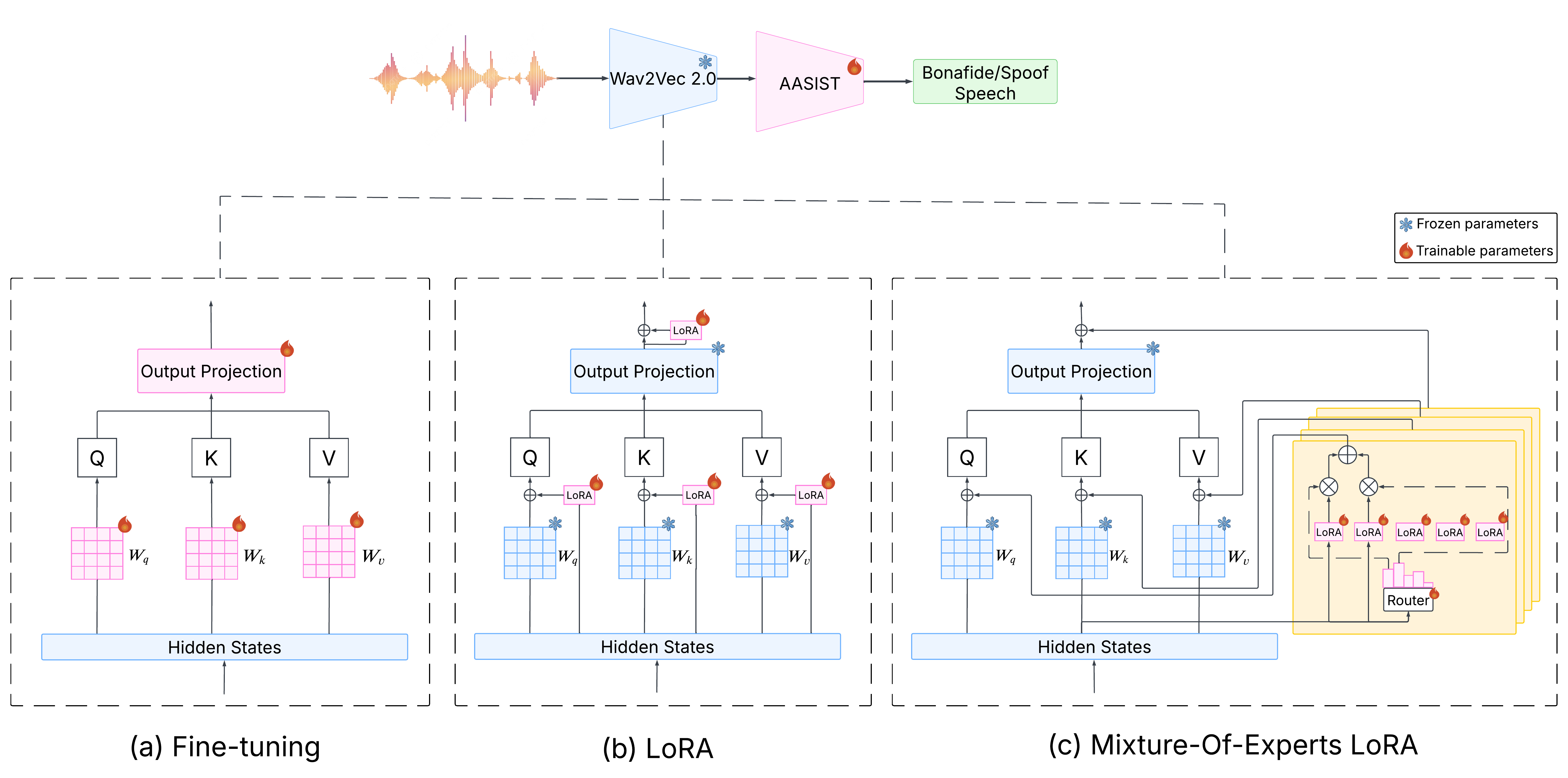}
    \caption{Overall scheme of the audio deepfake detection system (Wav2Vec2 + AASIST). We contrast the proposed (c) to baseline fine-tuning (a) and adapting only the LoRA \cite{laakkonen2025generalizablespeechdeepfakedetection} (b).}
    \label{fig:diagram}
\end{figure*}

Although Tak et al.~\cite{tak22_odyssey} demonstrated the effectiveness of combining Wav2Vec2 and AASIST for the ADD task, most existing approaches still struggle to adapt when confronted with unseen or evolving spoofing techniques \cite{muller22_interspeech}. Their reliance on fixed feature extraction or limited fine-tuning strategies often leads to domain overfitting, making it difficult to generalize beyond the conditions or attack types observed during training. Recent findings \cite{laakkonen2025generalizablespeechdeepfakedetection} indicate that LoRA-integrated models can surpass full fine-tuning in out-of-domain evaluations. Motivated by these results, we propose a sparse mixture-of-LoRA-expert framework that builds on the strong Wav2Vec2 + AASIST baseline. By integrating multiple LoRA experts within the attention layers of Wav2Vec2, our method employs a sparsely gated mechanism that dynamically selects and combines the outputs of a subset of these experts. This design allows the model to specialize in different aspects of the audio signal and to adapt to a wide range of spoofing cues. As a result, our framework improves generalization to out-of-domain attacks by leveraging both the parameter efficiency of LoRA and the adaptability of MoE.

The remainder of this paper is organized as follows. Section~2 details our proposed approach, introducing the underlying Wav2Vec2 + AASIST baseline and outlining how Mixture-of-LoRA Experts is integrated into the attention layers to improve out-of-domain generalization. In Section~3, we describe the experimental setup, including descriptions of the datasets used for both in-domain and out-of-domain evaluations, as well as training protocols and the evaluation metrics used for comparing performance across multiple datasets. Section~4 presents the results and discussion, comparing our approach with the baseline systems and conducting ablation studies to highlight the significance of each component in improving generalization performance. Finally, Section~5 concludes the paper by summarizing our key findings.

\section{Mixture-of-LoRA Experts}

Recent advancements in large language models have led to efficient techniques for scalability and generalization. Among them, the mixture of experts (MoE)~\cite{jacobs1991} and low-rank adaptation (LoRA)~\cite{hu2022lora} have gained popularity. Originally introduced in \cite{jacobs1991}, MoE has been widely used in speech processing \cite{SpeechMoE2021}, natural language understanding \cite{Fedus2021Switch}, and other applications. Specifically, it was explored for speech deepfake detection in~\cite{Negroni2024}.

Initially, low-rank adaptation (LoRA) was designed to efficiently fine-tune large language models~\cite{hu2022lora}. In Fig.~\ref{fig:diagram}, we see how LoRA can be applied to a transformer-based neural model. That specific model was used in~\cite{laakkonen2025generalizablespeechdeepfakedetection} for generalizable audio deepfake detection.

In the case that one LoRA is not enough, one can add more, where each one is an \emph{expert}. Then, a routing mechanism is needed to select an appropriate LoRA expert or subset of experts for a given input. This system is called the MoE-LoRA technique and has been applied in the context of large language models (LLM), AdaMoLE~\cite{liu2024adamole}. In this work, we investigate the fusion of MoE-LoRA for potential improvements in audio deepfake detection.

\textbf{Mixture-of-Experts.}
The mixture of experts (MoE)~\cite{ShazeerMMDLHD17} utilizes a framework of specialized models (experts) that collaboratively solve complex tasks based on the input features, dynamically selecting a subset of experts. Formally, a standard MoE module consists of a set of $N$ experts, $\{ E_i(\mathbf{x}) \}_{i=1}^{N}$, and a gating function $G_i(\mathbf{x})$ that dynamically coordinates the contribution of each expert. For each input $\mathbf{x}$, the gating function $G_i(\mathbf{x})$ has a trainable matrix $\text{W}_g$ to distribute the input $\mathbf{x}$ among the experts
\begin{equation}
    G_i(\mathbf{x}) = \text{Softmax}(\text{W}_g \, \mathbf{x} + \epsilon)_i,
\end{equation}
where Gaussian noise $\epsilon \sim \mathcal{N}(\mu, \sigma^2 I)$ with learnable mean $\mu$ and variance $\sigma^2$ encourages an exploration--exploitation trade-off; it promotes load balancing and helps avoid collapsing to a single most probable expert over time.
Only the top-$k$ experts are selected $\mathcal{S}(\mathbf{x}) = \operatorname{TopK}\{G_i(\mathbf{x})\}$, i.e., \emph{sparse selection}. If $k=N$, it is a \emph{dense} MoE variation, which is also explored in experiments. The output from the MoE layer is a weighted sum of the top-$k$ experts
\begin{equation} \label{eq:moe_output}
    \mathbf{y} = \sum_{i \in \mathcal{S}(\mathbf{x})} G_i(\mathbf{x}) \, E_i(\mathbf{x}).
\end{equation}

\textbf{Low-Rank Adapters.} The primary idea behind LoRA is to reduce the number of parameters needed for fine-tuning by approximating weight updates as low-rank matrices rather than updating the entire model's parameters. 
The general weight update in a neural network is defined as
\begin{equation}
\text{W}' = \text{W}_0 + \Delta \text{W},
\end{equation}
where $\text{W}_0$ represents the pre-trained weights of the backbone model, and $\Delta \text{W}$ represents the change introduced by fine-tuning. In LoRA, $\Delta \text{W}$ is parameterized as the product of two low-rank matrices:
\begin{equation}
\Delta \text{W} = \text{A}\,\text{B},
\end{equation}
where $\text{A} \in \mathbb{R}^{d \times r}$ and $\text{B} \in \mathbb{R}^{r \times m}$ are low-rank matrices with rank $r$, much smaller than the dimensions of $\text{W}_0$: $r \ll d, m$. 
Then, the output $h$ of the linear layer of the backbone model with fine-tuned LoRA is
\begin{equation} \label{eq:lora_output}
    \mathbf{h} = \text{W}_0\,\mathbf{x} + \Delta \text{W}\,\mathbf{x} = \text{W}_0\,\mathbf{x} + \text{A}\,\text{B}\,\mathbf{x}.
\end{equation}
This low-rank approximation drastically reduces the number of parameters that need to be learned, improving both the efficiency and flexibility of the fine-tuning process. LoRAs are typically added as side modules to the attention weights or feed-forward layers in the transformer. This allows the pre-trained model to retain its general knowledge while adapting to specific task requirements with minimal computational overhead. 

An additional benefit of LoRAs is that the $\text{A}$ and $\text{B}$ matrices can be stored separately from the backbone model. If the fine-tuning dataset is partitioned into segments, we can even train a separate set of $\text{A}$ and $\text{B}$ matrices for each segment. This idea then naturally leads to our contribution to the MoE-LoRA.

\textbf{MoE-LoRA.} The fusion of these approaches, termed MoE-LoRA, aims to enhance model efficiency and performance further; see Fig.~\ref{fig:diagram}. MoE enables the dynamic selection of experts, where specific LoRA experts detect different types of deepfake artifacts. Combining (\ref{eq:moe_output}) and (\ref{eq:lora_output}), the fusion output is 
\begin{equation}
    \mathbf{h} = \text{W}_0\,\mathbf{x} + \sum_{i \in \mathcal{S}(\mathbf{x})} G_i(\mathbf{x}) \, \big( \text{A}_i\,\text{B}_i\,\mathbf{x} \big),
\end{equation}
where each pair $(\text{A}_i, \text{B}_i)$ corresponds to a LoRA expert. We incorporate MoE-LoRA modules in each layer of the Wav2Vec2 backbone to explore the contribution of features in each layer.

\begin{figure}[!htb]
    \centering
    \includegraphics[width=\linewidth]{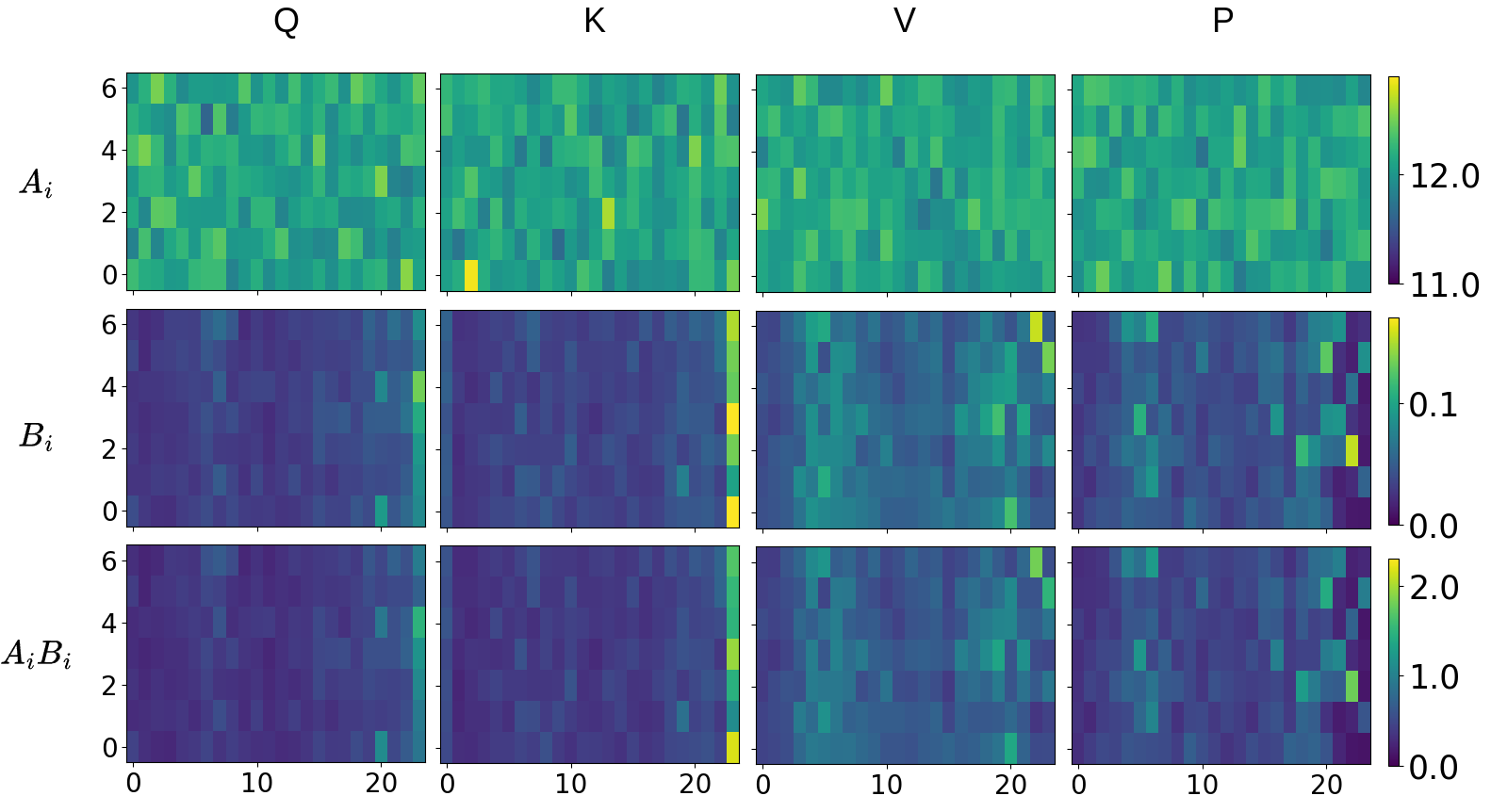}
    \caption{The maximal singular values of trained MoE-LoRA experts across 24 layers of the Wav2Vec2 backbone; the case of seven experts, top-$k$ is 7 (dense selection), and rank 8. Experts are indexed on the $y$-axis and backbone layers on the $x$-axis.}
    \label{fig:lora_expert_weights}
\end{figure}

In Fig.~\ref{fig:lora_expert_weights}, we can see a visualization of fine-tuned MoE-LoRA experts. The maximal singular value of each backbone layer--LoRA expert pair is denoted in the corresponding matrix entry. We observe that, for $\text{Q}$ and $\text{K}$ transformer matrices, only the last layers are significantly adapted. 
On the other hand, $\text{V}$ and $\text{P}$ (multi-head attention output projection) matrices see activity throughout the backbone layers.

\section{Experimental Setup}
\textbf{Datasets and evaluation metric:} We utilize the ASVspoof 2019~\cite{todisco2019asvspoof} Logical Access (LA) dataset for both training and validation, using its official training and development partitions. To assess the generalizability of our proposed method, we evaluate the models on several datasets:
\begin{itemize}
    \item \textbf{ASVspoof 2019 LA (evaluation split)}~\cite{todisco2019asvspoof}: The official evaluation partition from the same 2019 challenge is used to test performance consistency relative to the training domain.
    \item \textbf{ASVspoof 2021 LA and DF}~\cite{yamagishi2021asvspoof}: This comprises Logical Access (LA) and Deepfake (DF) attacks, offering a more diverse range of synthetic speech generation techniques.
    \item \textbf{ASVspoof 5 LA}~\cite{wang2024asvspoof}: A recently released, crowd-sourced dataset of $\sim$2{,}000 speakers recorded in diverse acoustic conditions, featuring 32 attack algorithms (including adversarial attacks).
    \item \textbf{In-The-Wild}~\cite{muller22_interspeech}: A curated 37.9-hour dataset of real and clearly faked audio featuring celebrities and politicians under varying conditions.
    \item \textbf{FakeAVCeleb}~\cite{khalid2021fakeavceleb}: A deepfake dataset derived from 500 celebrity videos in VoxCeleb2~\cite{chung18b_interspeech}; only the extracted audio is used.
\end{itemize}

\begin{table*}[!htbp]
\centering
\caption{Comparison of models trained with a single LoRA per layer vs. models trained with a mixture of LoRA experts (MoE). Performance is reported in terms of EER (\%), where bolded numbers are the best in each column and underlined are the second best. Sparse models use top-$k$=2, while dense models have top-$k$ equal to the number of experts.}
\setlength{\tabcolsep}{1mm}
\renewcommand{\arraystretch}{1.0}

\resizebox{\textwidth}{!}{
\begin{tabular}{lccc ccccccc}
\toprule
\multirow{2}{*}{Model} & \multirow{2}{*}{\shortstack{Trainable\\ Params.}} & \multirow{2}{*}{\shortstack{MoE\\ Experts}} & \multirow{2}{*}{\shortstack{LoRA\\ Rank}}  
& \multicolumn{7}{c}{Performance (EER \%)} \\
\cmidrule(lr){5-11}
& & & & ASV19:LA & ASV21:LA & ASV21:DF & In-The-Wild & FakeAVCeleb & ASV5 & Avg. \\
\midrule
Wav2Vec-AASIST  & 317.8M & --  & --  & 0.28 & 5.84 & 5.29 & 14.03 & 7.98 & 23.88 & 8.55 \\
Wav2Vec-AASIST* & 447K   & --  & --  & 0.36 & 4.29 & 7.97 & 19.41 & 4.84 & 17.14 & 9.00 \\
\midrule
\multirow{2}{*}{LoRA} 
& 1.23M & --  & 4  & 0.41 & 10.50 & 4.37 & 10.69 & 9.17 & 25.97 & 10.18 \\
& 2.02M & --  & 8  & 0.61 & 5.50 & 5.02 & 13.15 & 3.97 & 21.05 & 8.22 \\
\midrule
\multirow{6}{*}{Sparse MoE} 
& 3.40M  & 3  & 4  & 0.71 & 6.54 & 5.86 & 13.11 & 8.46 & 19.91 & 9.10 \\
& 5.36M  & 5  & 4  & 0.50 & 5.33 & 3.89 & 11.32 & \underline{1.81} & \underline{17.19} & \underline{6.67} \\
& 7.33M  & 7  & 4  & 0.48 & 4.48 & 5.84 & 14.72 & 2.95 & 19.11 & 7.93 \\
& 5.76M  & 3  & 8  & \textbf{0.26} & 5.73 & 6.81 & 9.75 & 6.96 & 22.39 & 8.65 \\
& 9.30M  & 5  & 8  & 0.34 & 6.06 & 4.69 & 10.59 & 8.77 & 23.66 & 9.02 \\
& 12.83M & 7  & 8  & 0.35 & 5.19 & 5.63 & \textbf{8.38} & 3.71 & 21.65 & 7.49 \\
\midrule
\multirow{6}{*}{Dense MoE} 
& 3.40M  & 3  & 4  & 0.38 & 5.95 & 6.78 & 10.60 & 6.30 & 22.49 & 8.75 \\
& 5.36M  & 5  & 4  & 0.42 & 6.37 & 4.18 & 9.11 & 4.05 & 20.80 & 7.42 \\
& 7.33M  & 7  & 4  & \textbf{0.26} & \textbf{3.70} & 4.01 & 15.59 & 1.96 & 18.41 & 7.31 \\
& 5.76M  & 3  & 8  & 0.29 & \underline{4.24} & \underline{3.70} & 9.77 & \textbf{1.77} & \textbf{16.75} & \textbf{6.08} \\
& 9.30M  & 5  & 8  & \underline{0.27} & 4.57 & 4.16 & 11.89 & 3.50 & 20.12 & 7.42 \\
& 12.83M & 7  & 8  & 0.69 & 5.35 & \textbf{3.25} & \underline{9.06} & 5.02 & 19.73 & 7.21 \\
\bottomrule
\end{tabular}
}
\label{tab:lora_vs_moe_comparison}
\end{table*}

\textbf{Baseline models.} For our primary baseline, we employ the Wav2Vec2 + AASIST system, inspired by previous advancements in speech deepfake detection \cite{tak22_odyssey, kukanov2024meta}. Specifically, we utilize Wav2Vec2 XLSR-53 (output dimension 1024) as the front end, coupled with AASIST---a spectrotemporal graph attention network---serving as the back-end classifier. In the fully fine-tuned variant, all parameters in both the SSL front end and AASIST are trainable. In contrast, we define Wav2Vec2 + AASIST* as a partially fine-tuned baseline, where the Wav2Vec2 front end remains frozen, and only AASIST is updated during training.

\textbf{LoRA models.} To explore parameter-efficient adaptations, we integrate Low-Rank Adapters (LoRA) into the Wav2Vec2 encoder's self-attention modules. In these models, only the LoRA parameters and the AASIST back end are trainable, while the rest of Wav2Vec2 remains frozen. For each self-attention block, LoRA matrices are inserted at the query, key, value, and output projections. We study single-LoRA configurations with rank $r\in\{4,8\}$.

\textbf{Mixture-of-LoRA-Experts (MoE-LoRA) models.} We extend the single-LoRA approach by introducing a mixture-of-experts mechanism within each self-attention block. Each block contains a set of LoRA experts---with ranks $r\in\{4,8\}$ and a gating router---and we vary the number of experts among $\{3,5,7\}$. During forward propagation, a sparse gating strategy selects the top-$k$ experts (with $k\in\{2,\text{num\_experts}\}$), providing a sparse or dense combination of experts. In MoE-LoRA models, the trainable parameters include the router parameters, the LoRA expert parameters, and AASIST. 

\textbf{Training strategy.} All variants are trained using the AdamW optimizer~\cite{loshchilov2019decoupledweightdecayregularization} with a cyclic learning-rate scheduler that varies the learning rate with a minimum of $1 \times 10^{-7}$ and a maximum of $1 \times 10^{-5}$ per cycle. The models are optimized to minimize the negative log-likelihood loss over two-class (bonafide vs. spoof) log-softmax outputs. Models are validated on the ASVspoof 2019 LA development set. Training terminates if no improvement is detected for a fixed number of epochs (10), retaining the best checkpoint.

\section{Results}

\begin{figure}[!htb]
    \centering
    \includegraphics[width=\linewidth]{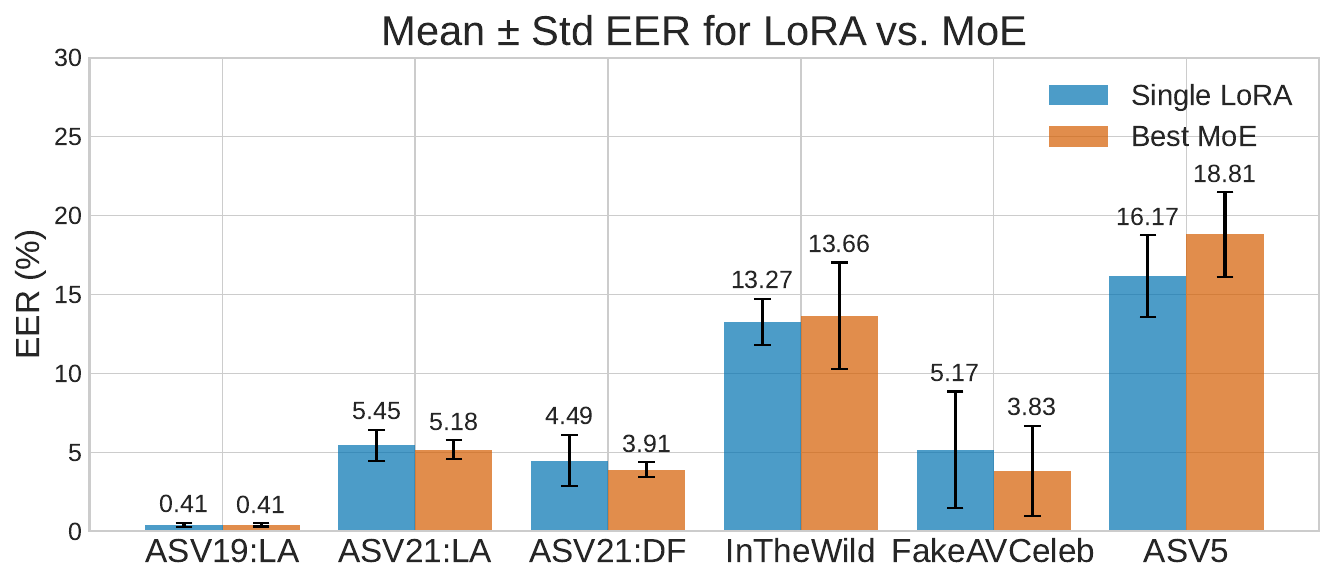}
    \caption{Mean $\pm$ std EER (\%) for LoRA and MoE models across five different seeds. The LoRA models use rank 8, while the MoE models have three experts, are dense (top-$k$=3), and also use rank 8. Error bars represent standard deviation across seeds.}
    \label{fig:barplot}
\end{figure}

To investigate the stability of our LoRA-based models, we evaluated both single-LoRA (rank=8) and MoE-LoRA (3 experts, top-$k$=3, rank=8) configurations under five different random seeds. Fig.~\ref{fig:barplot} summarizes the equal error rates (EER) for six test sets, namely ASVspoof 2019 LA, ASVspoof 2021 LA, ASVspoof 2021 DF, In-The-Wild, FakeAVCeleb, and ASVspoof~5. The single-LoRA setup has a notably high standard deviation on the FakeAVCeleb set (3.69\%), highlighting the impact of random initialization on performance, especially in challenging out-of-domain conditions. By contrast, the MoE-LoRA variant often achieved slightly lower average EER values than single-LoRA, though it also exhibited variability across seeds (for instance, standard deviations reached 3.37\% on the In-The-Wild corpus). On the ASVspoof~5 corpus, the single-LoRA configuration achieved an average EER of 16.17\% $\pm$ 2.60\%, while the MoE-LoRA model obtained 18.81\% $\pm$ 2.68\%, indicating no gain from the dense three-expert setup on this crowd-sourced, highly heterogeneous dataset. The variability in the evaluation EERs suggests that expert selection and gating can be sensitive to initialization, emphasizing the importance of aggregating or repeating trials when comparing approaches.

Table~\ref{tab:lora_vs_moe_comparison} compares the performance of three model types: fully or partially fine-tuned baselines (Wav2Vec2 + AASIST), single Low-Rank Adaptation (LoRA) configurations with varying ranks, and Mixture-of-LoRA-Experts (MoE-LoRA) variants. The fully fine-tuned Wav2Vec2 + AASIST baseline achieves an average EER of 8.55\% across all test sets; the partially fine-tuned version (frozen Wav2Vec2) yields a slightly higher EER of 9.00\%. While superior overall, the fully fine-tuned model struggles with out-of-domain generalization, achieving an EER exceeding 19\% on challenging datasets like ASVspoof~5.

Replacing full fine-tuning with a single LoRA layer within Wav2Vec2 demonstrates the effectiveness of parameter-efficient adaptation. A rank-4 LoRA narrows the performance gap. A rank-8 LoRA, however, achieves an average EER of 8.22\%, outperforming the partially fine-tuned approach and nearing the fully fine-tuned baseline. This highlights LoRA's ability to effectively adapt the model while keeping most Wav2Vec2 parameters frozen.

The MoE-LoRA framework further improves detection accuracy by utilizing multiple LoRA experts, each potentially specializing in different signal aspects. Each self-attention layer can employ either a sparse set of experts (a subset active at a time) or a dense set (all contributing). Sparse gating reduces computational load; dense gating can, in some cases, yield better performance. Notably, a dense MoE-LoRA configuration with three rank-8 experts achieves an average EER of 6.08\%, significantly outperforming both single-LoRA and the fine-tuned baselines. Adding more than three experts offers diminishing returns, with only marginal gains relative to the increased computational cost.

In summary, MoE-LoRA offers a compelling balance of flexibility and efficiency. By allowing expert specialization in detecting diverse spoofing cues, it achieves significantly lower error rates than single-LoRA or baseline fine-tuning. These results suggest that increased model capacity, combined with strategic gating and parameter-efficient adaptation, can markedly improve generalization to unseen deepfake attacks without a large increase in model size or computational demands.

\section{Conclusions}
Detecting audio deepfakes is challenging, as new generation techniques constantly outpace detection systems. We aim to improve the adaptability of audio foundation models to address this challenge. While Wav2Vec2 excels at audio representation learning, fine-tuned versions struggle with new deepfake types. Our results show a fully fine-tuned model averaging an 8.55\% EER across several out-of-domain sets. LoRA provides a partial solution, achieving 8.22\% EER. For significant generalization gains, we introduce a second representation-learning layer: a Mixture-of-Experts (MoE) approach. Combining multiple LoRA experts with strategic routing lowers the EER to 6.08\%. MoE-LoRA's adaptability offers a practical path towards reliable deepfake detection.

\section*{Acknowledgment}
This work was supported by the Finnish Doctoral Program Network in Artificial Intelligence, AI-DOC (decision number VN/3137/2024-OKM-6). The authors also wish to acknowledge CSC -- IT Center for Science, Finland, for computational resources. Additionally, Ville Hautam\"{a}ki thanks the Jane and Aatos Erkko Foundation for partial funding.

\bibliographystyle{IEEEtran}
\bibliography{Latex-2025/arxiv_ieee}

\end{document}